\RequirePackage{fixltx2e}
\documentclass[nofootinbib,reprint,showpacs,
  prc,aps,10pt,
  altaffilletter,floatfix]{revtex4-1}
\usepackage{graphicx}
\usepackage{hyperref}
\usepackage{amsmath}
\usepackage{amssymb}
\usepackage[amsmath,thmmarks]{ntheorem}
\usepackage{bm}
\usepackage{cleveref}
\usepackage[caption=false]{subfig}
\usepackage{array}
\usepackage{ifthen}
\usepackage{booktabs}
\usepackage{textgreek}
\usepackage{multirow}
\usepackage{enumitem}
\usepackage{dcolumn}
\usepackage{xspace}

\usepackage{color}

\RequirePackage{lineno}

\begin{document}

\newcommand{\comment}[1]{} 

\newcommand{\nue}       {$\nu_{e}$\xspace}
\newcommand{\numu}      {$\nu_{\mu}$\xspace}
\newcommand{\nutau}     {$\nu_{\tau}$\xspace}
\newcommand{\nubar}     {$\overline{\nu}$\xspace}
\newcommand{\nuebar}    {$\overline{\nu}_{e}$\xspace}
\newcommand{\numubar}   {$\overline{\nu}_{\mu}$\xspace}
\newcommand{\nutaubar}  {$\overline{\nu}_{\tau}$\xspace}
\newcommand{\znbb}      {$0\nu\beta\beta$\xspace}
\newcommand{\tnbb}      {$2\nu\beta\beta$\xspace}
\newcommand{\otsx}      {$^{136}\mathrm{Xe}$\xspace}
\newcommand{\dd}        {\mathrm{d}}
\newcommand{\rpm}{\raisebox{.2ex}{$\scriptstyle\pm$}}
\newcommand{\red}       {\textcolor{red}}
\newcommand{\mcl}       {$M_{\mathrm{cl}}$\xspace}
\newcommand{\mmmcl}       {M_{\mathrm{cl}}}

\title{Measurement of neutron capture on \otsx}

\author{J. B. Albert}\email{joalbert@indiana.edu}
\author{S. J. Daugherty}
\author{T. N. Johnson}
\altaffiliation{Now at University of California, Davis, CA 95616, USA}
\author{T. O'Conner}
\author{L. J. Kaufman}
\affiliation{Physics Department and CEEM, Indiana University, Bloomington, 
Indiana 47405, USA}

\author{A. Couture}
\author{J. L. Ullmann}
\affiliation{Los Alamos National Laboratory, Los Alamos, New Mexico 87545, USA}

\author{M. Krti\v{c}ka}
\affiliation{Faculty of Mathematics and Physics, Charles University in Prague, 
V Hole\v{s}ovi\v{c}k\'{a}ch 2, CZ-180 00 Prague 8, Czech Republic}

\date{\today}

\begin{abstract}

$^{136}$Xe is a 0$\nu\beta\beta$ decay candidate isotope, and is used 
in multiple 
experiments searching for this hypothetical decay mode.
These experiments require precise information about neutron capture for 
their background 
characterization and minimization. Thermal and resonant neutron capture 
on $^{136}$Xe have 
been measured at the Detector for Advanced Neutron Capture Experiments 
(DANCE) at the Los 
Alamos Neutron Science Center. A neutron beam ranging from thermal energy 
to greater than 100 keV was incident on a gas cell filled
with isotopically pure $^{136}$Xe.  The relative neutron capture cross sections 
for neutrons at thermal energies and the first resonance at 2.154 keV 
have been measured, yielding a new absolute measurement of $0.238 \pm 0.019$~b 
for the thermal neutron capture
cross section.  Additionally, the $\gamma$ cascades
for captures at both energies have been measured, and cascade models have been 
developed which
may be used by $0\nu\beta\beta$ experiments using \otsx.

\end{abstract}

\maketitle


\section{\label{sec:intro}Introduction}

Neutrinoless double beta decay (\znbb) is a hypothetical 
lepton-number-violating decay mode.
Observation of \znbb would be a confirmation that neutrinos 
are Majorana particles;
i.e.,~there is no distinction between neutrinos and 
antineutrinos.  Nonobservation of this
process, combined with information about the absolute mass of 
neutrinos, may be used to
demonstrate that neutrinos are Dirac particles, in which case 
neutrinos and anti-neutrinos
have an intrinsic distinction.  As the nature of the neutrino is 
of considerable interest
for understanding the standard model, several experimental 
collaborations are running
or developing experiments to search for \znbb.

One of the most common isotopes to use for this search 
is \otsx~\cite{Auger:2012gs, Gando:2012zm, Martin-Albo:2015rhw}.
This isotope is ideal in many ways, including the large 
$Q$ value (2457.83~keV~\cite{Redshaw:2007un}),
ease of enrichment, and physical characteristics allowing for scaling 
to large detectors.
Due to the rarity of \znbb decays, a successful search requires extremely low 
radioactivity in detector materials to minimize backgrounds.
As a noble gas, \otsx can be highly purified, and detectors can be 
constructed with extremely
radiopure materials.  Techniques such as 
multiplicity discrimination~\cite{Auger:2012ar} can be
used to further reduce backgrounds due to $\gamma$ rays from 
radioactive decays.  One background
which cannot be reduced through these techniques is the 
$\beta$ decay of $^{137}\mathrm{Xe}$.
In a recent \znbb search by the EXO-200 
Collaboration~\cite{Albert:2014awa}, $^{137}\mathrm{Xe}$
$\beta$ decay was estimated to be responsible for 20\% of 
backgrounds in the \znbb
signal region of interest.

A separate study by the EXO-200 Collaboration~\cite{EXO200::2015wtc} found that 
$^{137}\mathrm{Xe}$ in the detector was overwhelmingly produced by 
$^{136}\mathrm{Xe}(n,\gamma)^{137}\mathrm{Xe}$ interactions 
with neutrons produced from
cosmic-ray muon interactions underground.  These neutrons typically thermalize
in the shielding around the xenon before capturing.  It is possible to 
reject a significant fraction
of this background by identifying the production of $^{137}\mathrm{Xe}$ 
and implementing a veto
to remove the subsequent decays (3.8~minute 
half-life~\cite{Carlson1969267}) from the dataset.

To better understand backgrounds and to facilitate the development 
of such a veto, we have studied the 
$^{136}\mathrm{Xe}(n,\gamma)^{137}\mathrm{Xe}$ interaction using the 
Detector for Advanced
Neutron Capture Experiments (DANCE).  The relative capture cross sections 
for thermal neutrons and neutrons at 2.154 keV, the first \otsx resonance, 
were measured, as well as the energies and
multiplicities of cascade $\gamma$'s for thermal and resonant captures.
This information
may be used by EXO-200 and other collaborations to improve the 
sensitivity of their
\znbb searches, and may also provide insight into the nuclear 
structure of $^{137}\mathrm{Xe}$.
Additionally, this measurement can yield a new absolute cross section for
thermal neutrons when combined with an external measurement of neutron
capture at resonance.

\section{\label{sec:exp_meth}Experimental Method}
\subsection{\label{dance}DANCE}
DANCE is located on Flight Path 14
at the Manuel Lujan Jr. Neutron Scattering Center at the Los Alamos 
Neutron Science Center. 
This flight path is exposed to neutrons that pass through a 
room-temperature water moderator.
The target sample, centered within the detector, is 20.25~m 
downstream of the moderator. 
Prompt $\gamma$ rays are measured from neutron capture using 160
BaF$_{2}$ crystals arranged spherically around the target, 
covering a solid angle of 
$\sim$3.5$\pi$ steradians.  Each crystal is 15~cm long, has a volume 
of 734~cm$^3$, and is
monitored by a photomultiplier tube (PMT).
BaF$_2$ crystals have fast timing resolution, which allows for precise neutron 
time-of-flight measurement,
and the segmentation is ideal for measurement of $\gamma$-cascade multiplicity.
The space between the evacuated beam pipe and the inner surfaces 
of the crystals (at 16.5~cm
radius~\cite{Ullmann:2014roa}) is filled with a $^{6}\mathrm{LiH}$ 
shell to reduce the rate of scattered neutrons
capturing on the BaF$_{2}$ crystals.
Further information on the detector can be 
found in Ref.~\cite{Reifarth:2013xny}.

\subsection{\label{daq}Data Acquisition}
Neutrons were incident on a 3~cm thick sample of 99.9\% pure, 
gaseous $^{136}\mathrm{Xe}$ pressurized to an average
of 26 psi. The xenon gas was contained in an aluminum cell with 
2.9~cm diameter, 0.003~inch thick kapton windows 
allowing the neutron beam to pass through.  
Data were also taken with the same cell evacuated, allowing for determination 
of the beam and target-related backgrounds.
As the beam diameter was smaller than 2~cm at the target, the 
full flux of neutrons was incident
upon xenon.

The data was collected by two digitizers each recording a 
256~$\mu$s long window.  These time windows were set
to a delay relative to the initial neutron beam trigger to 
select specific neutron energies
based on time of flight. The first time window was set to 
look at the high neutron 
energy events, including the
2.154 keV $^{136}\mathrm{Xe}$ capture resonance, while the second 
was delayed by 9.15 ms
to look at the thermal neutron energy range
of 0.0243 to 0.0256~eV. Within each of these time windows, all signals 
from the PMTs mounted to the crystals were recorded.
The energy windows used in analysis are shown in Fig.~\ref{fig:flux}.

\subsection{\label{flux}Neutron Flux Determination}

Located downstream of the sample location are three neutron monitors that 
are used to measure the neutron flux as a function of energy.  These 
monitors use the $^{6}\mathrm{Li}(n,\alpha t)$
reaction, the $^{235}\mathrm{U}(n,f)$ reaction, and 
the $^{3}\mathrm{He}(n,p)$ reaction.
As the beam diameter is smaller than both the xenon target and 
the beam monitor, we measure
the total neutron rate per beam spill as a function of time of flight 
(which is converted to neutron energy).
The $^{6}\mathrm{Li}(n,\alpha t)$ monitor has
good performance at both thermal and resonant energies, so it was 
used for this measurement.
The $^{3}\mathrm{He}(n,p)$ and $^{235}\mathrm{U}(n,f)$ monitors were 
used for cross-checks,
and showed good agreement for the measured flux shape.

\begin{figure}[htb]
\includegraphics[width=\columnwidth]{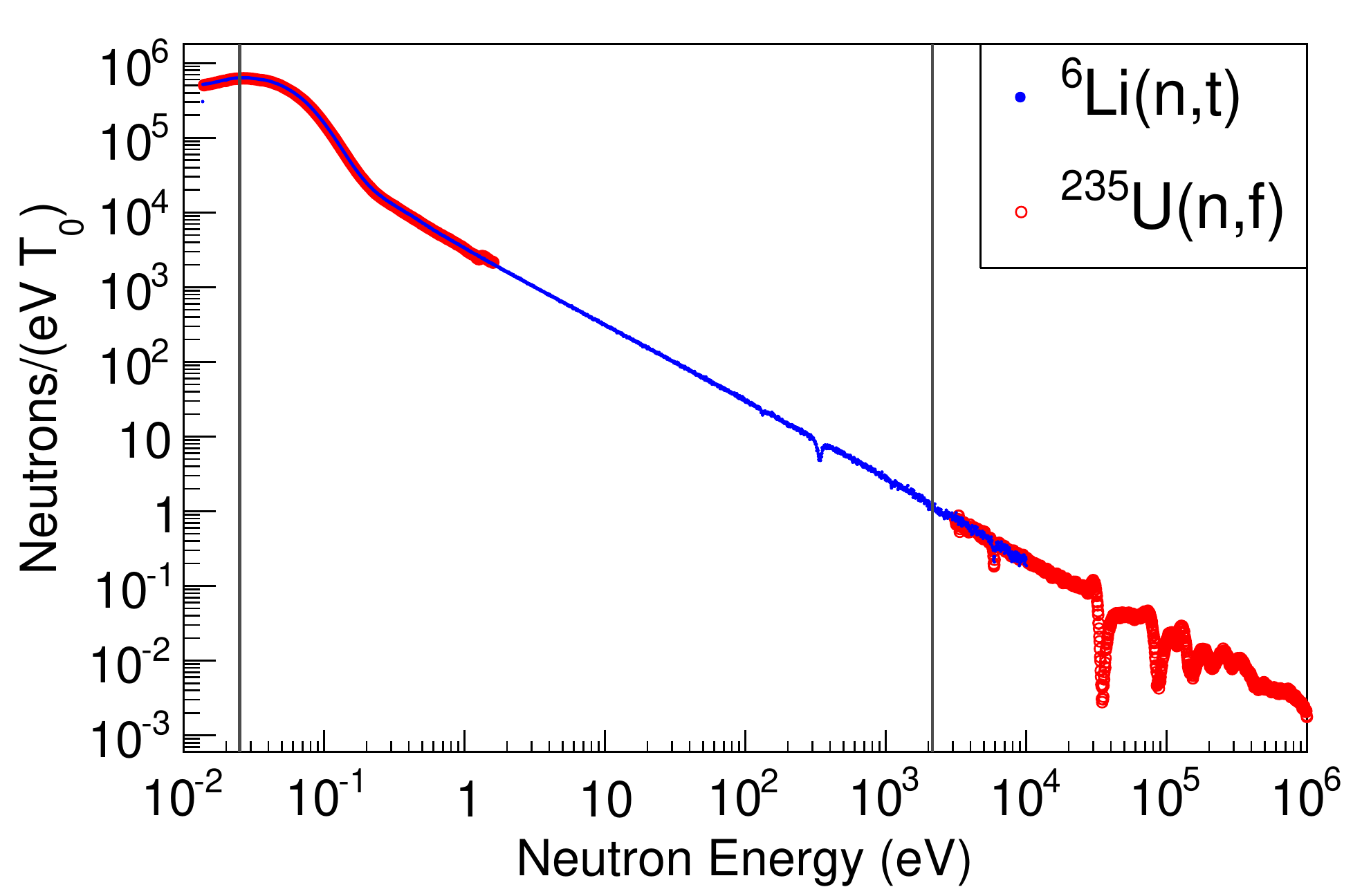}
\caption{(Color online) Neutron flux measured by the $^{6}\mathrm{Li}$ 
and $^{235}\mathrm{U}$ neutron monitors. The flux is integrated over the
regions indicated by the dark grey bars for calculation of our thermal 
and resonance cross sections.  The gap
in flux measurement using the $^{235}\mathrm{U}$ monitor is due to an 
energy region with
resonances that make the flux difficult to evaluate.}
\label{fig:flux}
\end{figure}

The neutron rate was determined by using a surface barrier Si detector 
to count $^{6}\mathrm{Li}(n,\alpha t)$ interactions in a
$^{6}\mathrm{LiF}$ layer deposited on a thin kapton film.
The number of interactions were converted to a flux measurement 
using knowledge of the 
beam and detector geometry and the known 
cross section~\cite{ENDF6Li6:1991:xxx} for this interaction.
The measured flux as a function of incident neutron energy is 
shown in Fig.~\ref{fig:flux}.
As only the ratio of fluxes at different energies is necessary 
for this analysis, 
uncertainties due to the absolute calibration of the neutron 
monitors are negligible.

\section{\label{sec:data_ana}Data Analysis}

\subsection{Event Reconstruction}
\label{sec:evt_recon}

After a applying a timing calibration, all PMT signals occurring within 
a 20~ns window are grouped
together as a single event.  
Only crystals with measured energy above the threshold of 250~keV are counted.
We determined, based on measured event rates and Poisson statistics, 
that the probability for 
two or more neutron-induced
events to overlap within a single 20~ns coincidence window is less 
than 1\% at the capture resonance energy,
and less than 0.1\% at thermal neutron energies, so pile-up effects 
are negligible.  This coincidence window is wide enough that uncertainties 
in relative timings for each PMT do not
significantly affect efficiency.

Scintillation light
in the BaF$_{2}$ crystals has a fast ($\sim$0.6~ns) and slow 
($\sim$0.6~$\mu$s) component.
The ratio of fast to slow scintillation light can be used to 
discriminate between 
$\alpha$-induced signals and those from $\beta$ decay or $\gamma$ rays.  
This discrimination allows
for a near perfect suppression of $\alpha$ backgrounds to neutron 
capture signals.  
The $\alpha$ decay signals were collected and used for the energy 
calibration of the BaF$_{2}$ crystals.

The remaining events with $\gamma$-like fast/slow ratios were 
analyzed for neutron capture studies.
  Often, $\gamma$ rays from neutron captures will Compton scatter and 
  deposit energy in
multiple adjacent crystals.  Thus, to reconstruct the $\gamma$ ray 
multiplicity and the full
energy of each $\gamma$ ray, a clustering algorithm was implemented.
Adjacent crystals recording signals in a single 20~ns coincidence 
window are grouped together as a 
cluster, and it has been found that these clusters correspond well 
with individual $\gamma$ rays.  
The reconstructed number of clusters (\mcl), individual cluster 
energies ($E_{\mathrm{cl}}$), sum of all cluster energies
($E_{\Sigma}$), and neutron energy ($E_n$, measured 
from time of flight) are used
in this analysis.

\subsection{Background Subtraction}
\label{sec:bkg_sub}

\begin{figure*}[hbt]

\includegraphics[width=\textwidth]{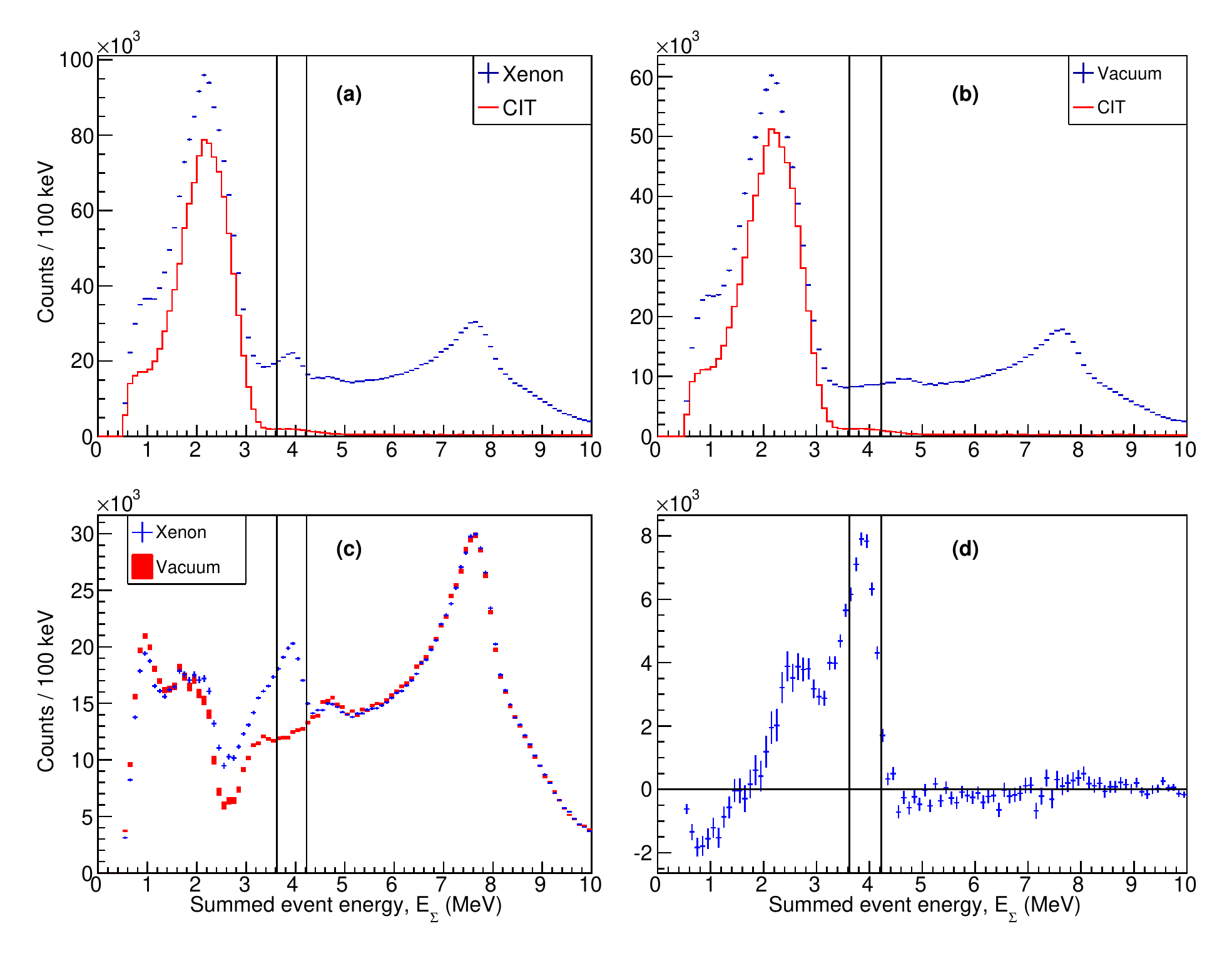}

\caption{ (Color online)
(a) Summed event energy for pressurized xenon data at the 25 meV 
neutron energy window 
and for beam-off data (also called constant in time, CIT), 
including cluster multiplicities
2 through 5. The beam-off data have been scaled by livetime.  
(b) Same as (a), but with 
evacuated target data rather than pressurized xenon data.  
(c) Summed event energy
for pressurized xenon data and evacuated target data both at the 
25 meV neutron energy window after the CIT backgrounds have been
subtracted. The evacuated target spectrum has been scaled so that the 
counts in the 6 to 9 MeV region match the pressurized xenon data.  
(d) Summed event energy spectrum after the evacuated
target data has been subtracted from the pressurized xenon data.
For all panels, the $Q$~gate used in this analysis is indicated by 
black vertical lines.}
\label{fig:bkg_sub}
\end{figure*}

\begin{figure*}[htb]
\includegraphics[width=\textwidth]{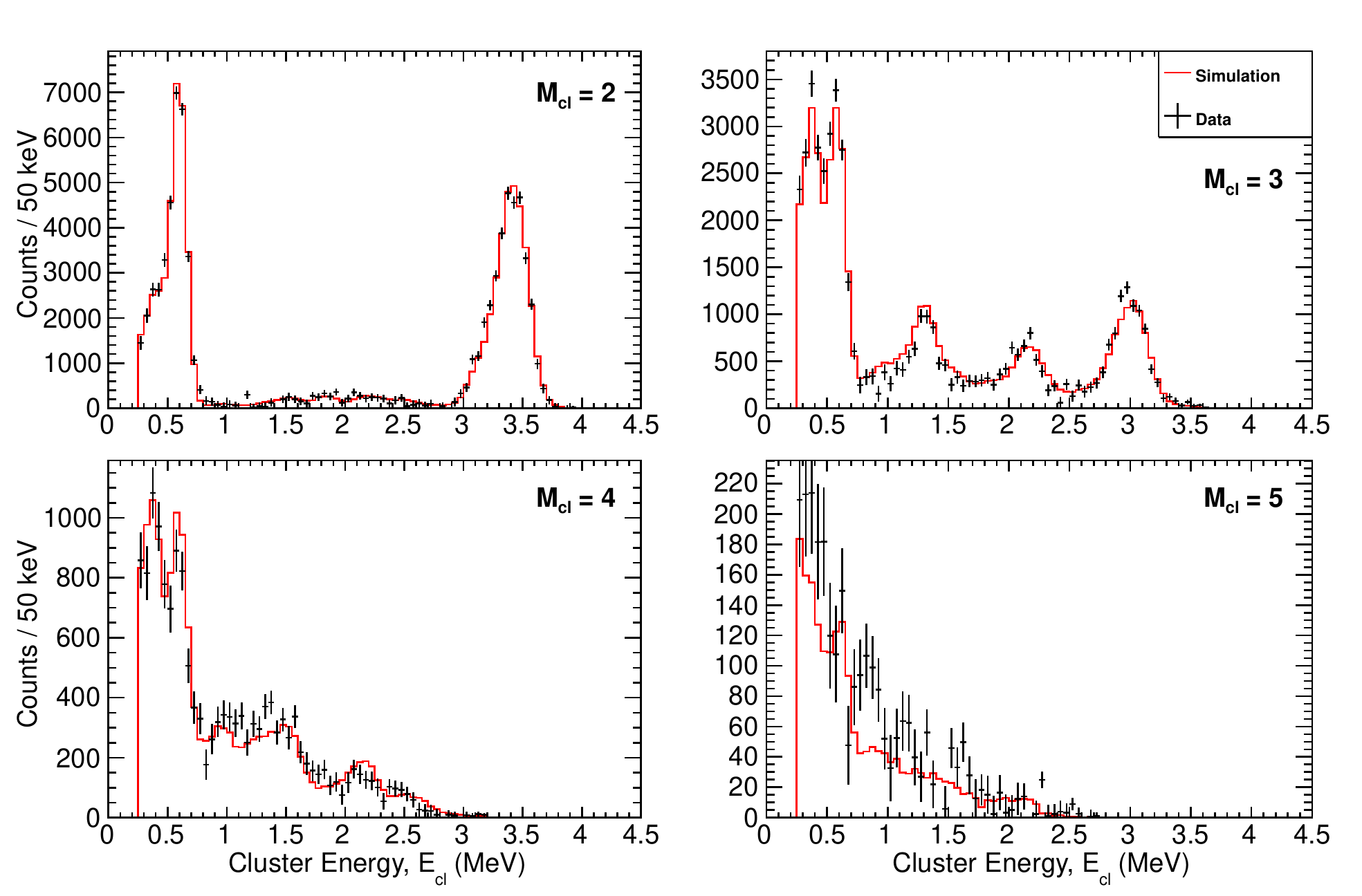}

\caption{(Color online) Measured and simulated MSC spectra from 
the $^{136}\mathrm{Xe}$ capture cascade in the 25 meV neutron energy window 
for cluster multiplicities 2 through 5.
The simulated spectra, shown in red, represent the DICEBOX 
realization best matching the
experimental data.}
\label{fig:thermal_msc}
\end{figure*}

\begin{figure*}[htb]
\includegraphics[width=\textwidth]{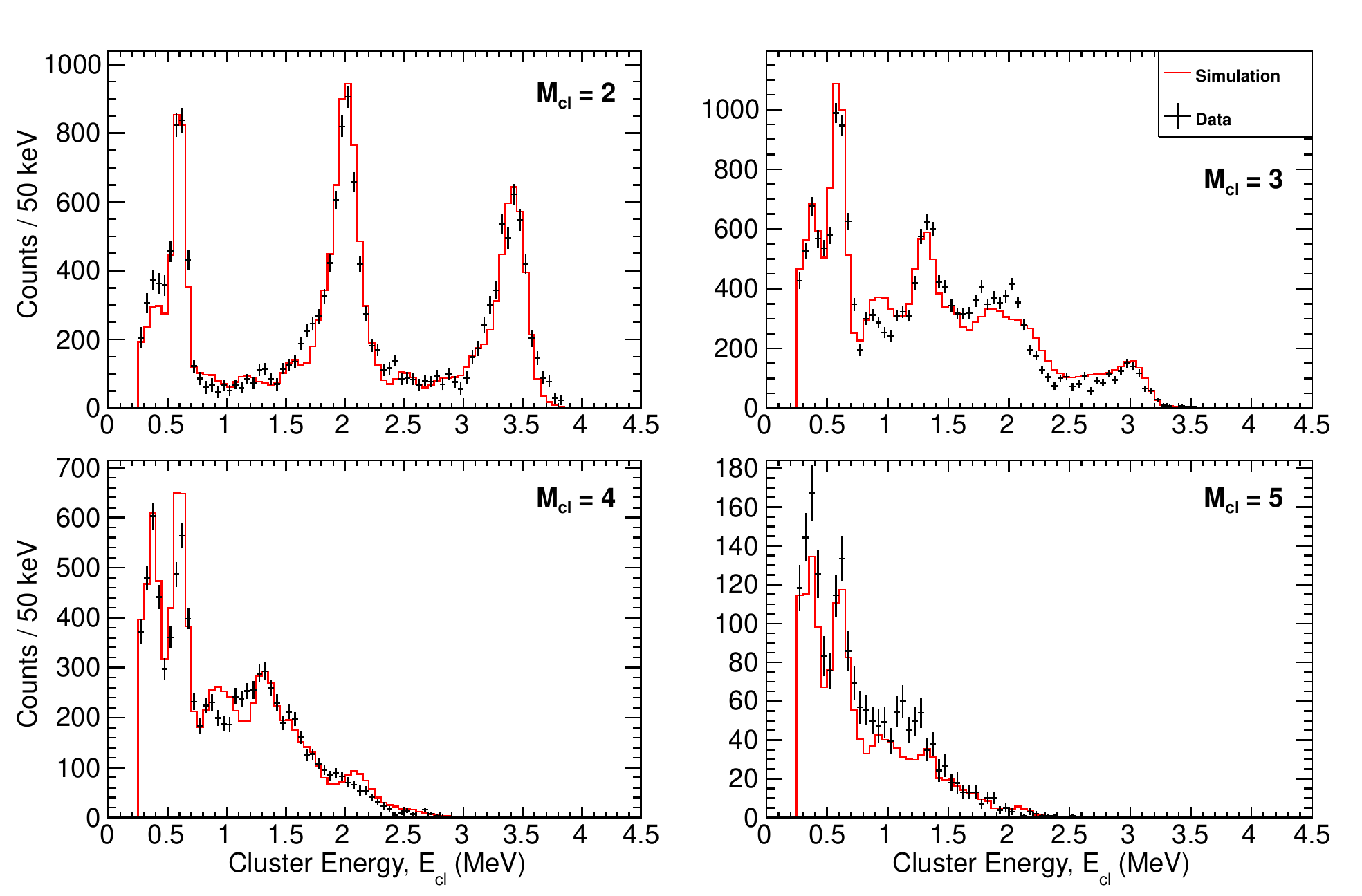}

\caption{(Color online) Measured and simulated MSC spectra from the 
$^{136}\mathrm{Xe}$ capture cascade at the 2.154~keV neutron energy 
resonance for cluster multiplicities 2 through 5.
The simulated spectra, shown in red, represent the DICEBOX 
realization best matching the
experimental data. }
\label{fig:resonant_msc}
\end{figure*}

Because radioactive $^{226}\mathrm{Ra}$ is a chemical homologue to 
barium, the crystals have some
radioactive contamination.  The decay chain from $^{226}\mathrm{Ra}$ 
includes several
$\alpha$ decays, as well as some decays with $\beta$'s 
and $\gamma$'s ($^{214}\mathrm{Pb}$ and $^{214}\mathrm{Bi}$ in particular). 
The $\alpha$ decays are easily rejected with the technique described in
Sec.~\ref{sec:evt_recon}, but the $\beta$ decays in the crystals 
(and from outside
the detector) produce a constant-in-time~(CIT) background to neutron capture.
This CIT background dominates single-cluster data, and some CIT events have 
$M_{\mathrm{cl}} \ge 2$ due to $\beta$ decays which are 
accompanied by $\gamma$ rays, 
producing a multicluster event.

Another background comes from
beam neutrons which may scatter off the xenon and capture on 
aluminum in the target vessel or
beam pipe, or on barium in the crystals.  Captures from scattered 
neutrons dominate
the data for $E_{\Sigma} > 3$~MeV.

To appropriately subtract these backgrounds, three separate datasets 
were used: pressurized xenon target with incident beam (pressurized xenon data),
evacuated target with incident beam (evacuated target data), and no target with 
no neutron beam (beam-off data). The evacuated target
data measure backgrounds due to scattered neutron capture 
(as neutrons may still scatter
off the kapton windows), and beam-off data measure the CIT backgrounds. 
While the scattered neutron capture backgrounds seen at different 
times of flight scale with the
number of scattered neutrons, CIT backgrounds
scale only with livetime, so these must be treated separately.

First, the CIT data were scaled to match the livetimes of both the 
pressurized xenon and evacuated target data,
and the CIT spectra were subtracted from the pressurized xenon and 
evacuated target spectra
at each multiplicity.
Panels (a) and (b) in Fig.~\ref{fig:bkg_sub} show the CIT background 
scaled to the pressurized xenon
and evacuated target data in the thermal neutron energy window.
Only a very small fraction of the CIT background 
has $E_{\Sigma} > 3.5~\mathrm{MeV}$.

Panel (c) of Fig.~\ref{fig:bkg_sub} shows the resultant 
spectra for the pressurized xenon
and evacuated targets after the CIT background subtraction. 
The evacuated target spectrum was scaled
to match the number of counts in the pressurized xenon spectrum 
in the 6 to 9 MeV 
$E_{\Sigma}$ range. This is well above the 4.025 MeV 
$^{136}\mathrm{Xe}$ neutron capture $Q$ value, 
so the events in this range are only due to scattered neutrons.  
In this way, CIT 
backgrounds and scattered neutron capture backgrounds are 
appropriately subtracted, as
seen in panel (d) of Fig.~\ref{fig:bkg_sub}, leaving a large peak 
at the $^{136}$Xe$(n,\gamma)$ $Q$ value
and an excess at lower energies due to Xe capture events where some 
fraction of the 
$\gamma$ cascade energy is lost.  While Fig.~\ref{fig:bkg_sub} 
illustrates the subtraction
process with $E_{\Sigma}$ spectra, the same procedure, with the same 
scale factors, is
 applied
to all relevant spectra, including those of individual cluster energies.

A valley in both the signal and background is apparent between 
2.2 and 3.2 MeV in  
panel (c) of Fig.~\ref{fig:bkg_sub}. This may be 
due to a small energy mis-calibration with the beam-off data, or 
imperfect background 
subtraction.
This valley is most apparent where the slope of the beam-off 
spectrum is steepest.  The beam-off
spectrum is small relative to \otsx capture and relatively flat in 
the region near
the \otsx $Q$ value, so any possible energy miscalibration would have a 
negligible effect on the
analysis.  The beam-on pressurized xenon and evacuated target data 
share the same
energy calibration.

As both the CIT and scattered neutron backgrounds largely come 
from $\gamma$ or $\beta$
emission inside a single crystal, the data for $\mmmcl=1$ are 
dominated by backgrounds.
Hence, this analysis largely uses only $\mmmcl > 1$ spectra.  There were 
almost no events
with $\mmmcl > 5$.  To minimize errors due to imperfect background 
subtraction, we
further restricted the analysis to events with 
$3.625~\mathrm{MeV} < E_{\Sigma} \le 4.225~\mathrm{MeV}$.  
This ``$Q$~gate'' optimizes 
the signal-to-background 
ratio and avoids most of the CIT backgrounds while still 
leaving good statistics.  One exception, where $\mmmcl = 1$
data was used, is discussed in Sec.~\ref{sec:cascade_modeling}.

One additional background source comes from $\gamma$ rays 
(mainly 2.2~MeV from capture on hydrogen in the neutron moderator)
that may travel down the beam pipe and pair-produce in the xenon, 
yielding a pair of 
0.511~MeV $\gamma$'s due to positron annihilation.  These signals 
are mainly found at short
time of flight, and have a total energy far below the $Q$~gate.
No subtraction of these beam
backgrounds was necessary, as they could not affect the analysis.

\subsection{Cascade Modeling}
\label{sec:cascade_modeling}

To optimize identification of the \otsx(n,$\gamma$) reaction in 
\znbb searches such
as EXO-200, the cascade from the capture to the ground state 
of $^{137}\mathrm{Xe}$ must be
known as precisely as possible. 
We use multistep cascade (MSC) spectra to evaluate cascade models.  We define
MSC spectra as the spectra
of $E_{\mathrm{cl}}$ at each cluster multiplicity ($M_{\mathrm{cl}} = 2-5$).
We compare the MSC spectra measured with the DANCE detector to predictions 
derived from simulations and candidate cascade models. 
We use a Geant4~\cite{Allison:2006ve, Agostinelli:2002hh} simulation 
which features the DANCE geometry and detector 
response~\cite{Jandel20071117} for cascades produced with the DICEBOX 
code~\cite{Becvar1998434} in a way similar to that in Ref.~\cite{Rusev88057602}.
For this analysis, we added the geometry of the aluminum target and 
pressurized xenon into 
the Geant4 simulations and assumed that the captures occur uniformly in 
the Xe target.

The DICEBOX code uses existing information on levels below a certain 
critical energy ($E_c = 2.65$ MeV in this analysis), 
including intensities of primary transition to these levels and 
subsequent transitions.
Individual levels above $E_c$ and $\gamma$ transitions from these 
levels are generated 
``randomly'' based on statistical models of nuclear level density and 
photon strength functions. 
Each set of levels and transitions is 
called a ``nuclear realization''~\cite{Becvar1998434}. Assuming the data 
for levels
below $E_c$ is accurate and complete, and given enough realizations, a 
model closely
matching the cascade found in nature should be achievable.  After 
the information on 
levels below $E_c$ was finalized,
100 nuclear realizations were simulated for thermal capture, and 200 
for resonant capture,
each with $10^5$ cascades.
The nuclear realization best describing the spectra was chosen based on 
the global $\chi^2$ agreement for all bins in MSC spectra for
 $M_{\mathrm{cl}}=2-5$.  The chosen realizations
were re-produced with $10^6$ cascades, for better statistics.
Fig.~\ref{fig:thermal_msc} shows the agreement of MSC spectra for the 
chosen nuclear realization with the experiment for the thermal neutron 
energy window.
There is only one common normalization factor for all multiplicities, so 
the good
agreement indicates an accurate multiplicity distribution.

The information on the decay scheme below $E_c$ was taken from 
ENSDF~\cite{Browne20072173}, largely based on the thermal neutron capture work 
by Prussin \textit{et al.}~\cite{Prussin:1977zz}. Transition intensities were 
slightly adjusted to improve the agreement between data and simulations. The 
changes
to the thermal capture cascade, from that described in Prussin's measurement, 
were relatively minor. 
On the other hand, no information on cascade transitions was available for decay 
of the 
2.154~keV resonance. The primary transitions from this resonance were initially 
based on
the thermal cascade model, but significant adjustments were made manually to 
reproduce the 
resonance MSC spectra. 

The neutron capture cascades for the thermal neutron window and 2.154 keV 
resonance window show 
significant differences, as seen in Figs.~\ref{fig:thermal_msc} 
and~\ref{fig:resonant_msc}. This is not surprising as the initial states 
are different:
the 2.154 keV resonance is a $p$-wave $3/2^{-}$ state, while thermal
neutrons ($s$-wave) produce a $1/2^{+}$ state.

The most visible difference is a strong two-step cascade seen in the middle of 
the
$M_{\mathrm{cl}}=2$ MSC spectrum for the resonance. Its presence indicates the 
existence of a $J=5/2$ state at 
$E \simeq 2$~MeV; this is the only spin which allows dipole transitions to 
connect 
the neutron capturing state (presumed to be 
$J^{\pi}=3/2^{-}$~\cite{mughabghab2006atlas}) with the ground state 
($J^{\pi}=7/2^{-}$). A level with this spin cannot be strongly populated in 
thermal 
neutron capture as it cannot be accessed via a dipole primary transition from 
the thermal capture
state ($J^{\pi}=1/2^{+}$). Several levels near 2~MeV excitation energy have been 
reported from studies of $\beta$ decay of the $7/2^+$ $^{137}\mathrm{I}$ ground 
state~\cite{Browne20072173}. 

Direct transition from the thermal capture state to the ground state would 
require an octopole transition (extremely suppressed), and has not been 
observed in
previous experiments~\cite{Prussin:1977zz}.  However, for capture at the 
2.154~keV
resonance, a direct transition to the ground state could be achieved with an
electric quadrupole transition, and would show up in the $\mmmcl = 1$ data 
as a peak
at $E_{\mathrm{cl}} = 4027$~keV.  As the $\mmmcl = 1$ data 
is background dominated and contains
important features missed with the usual $Q$~gate selection on $E_{\Sigma}$,
a separate study was performed to 
measure the possible intensity of this primary transition.  After background 
subtraction (as described in Sec.~\ref{sec:bkg_sub}), 
the expected peaks at $E_{\mathrm{cl}} = E_{\Sigma} = 4027$ and 
3424~keV were observed in the $M_{\mathrm{cl}}=1$ spectrum.  Peaks at lower 
energies were unusable due to large
unsubtracted backgrounds.  Simulations of the known capture cascade were 
performed
with varying intensities of the direct transition to the ground state until the 
two peaks
were well reproduced, although a relatively flat background spectrum of unknown 
origin remained. 
Based on this, we determined that resonant captures will transition
directly to the ground state $2.3\pm1.0\%$ of the time.  This  contribution to 
the decay was added to the
resonant cascade model for DICEBOX.  Because the $\mmmcl = 2-5$ data are not 
sensitive to this transition, and because the $\mmmcl = 1$ backgrounds are not 
fully
understood, a separate systematic uncertainty was included 
to account for this transition.

\begin{figure*}[htb]
\subfloat{
  \includegraphics[width=.49\linewidth]{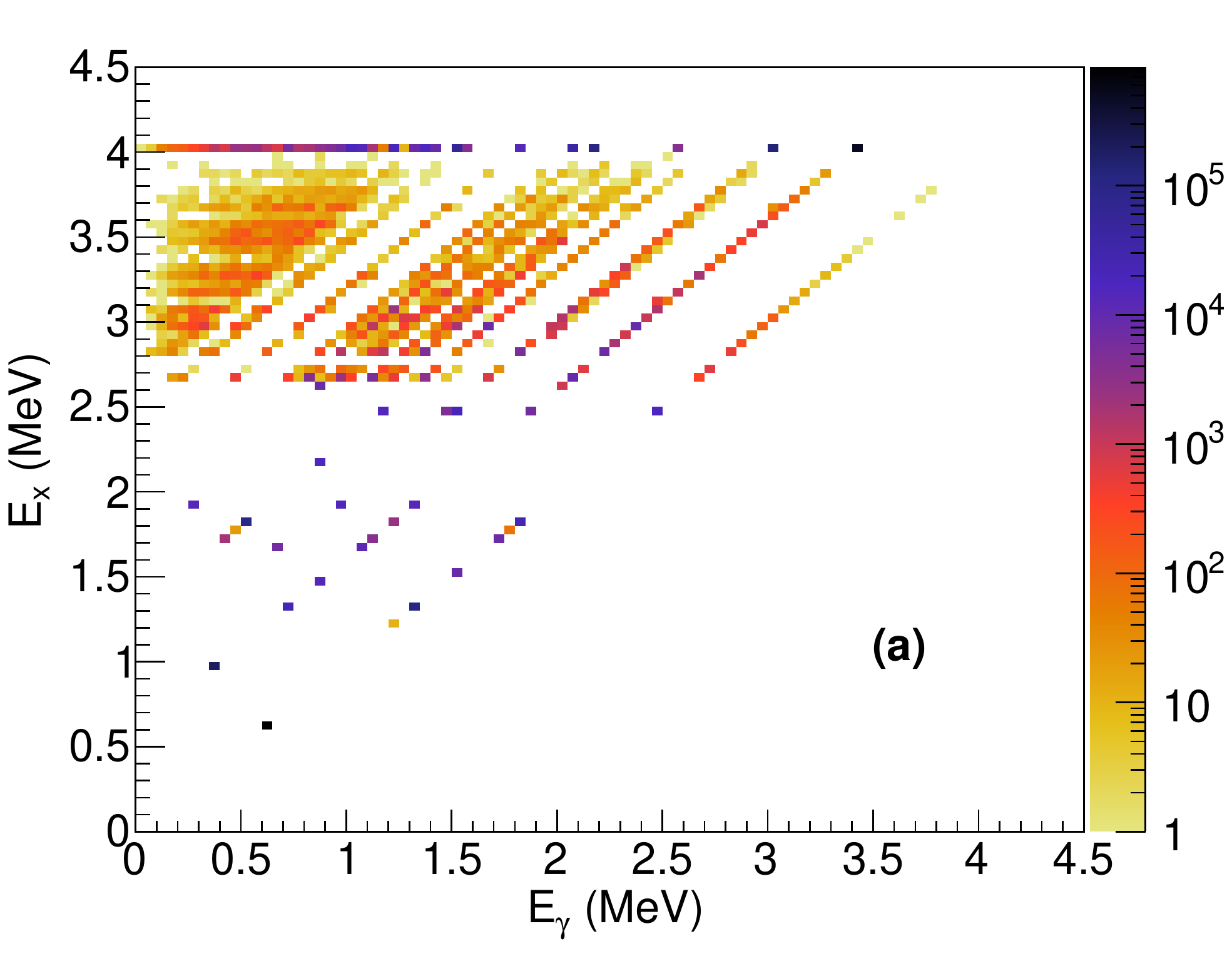}
}
\subfloat{
  \includegraphics[width=.49\linewidth]{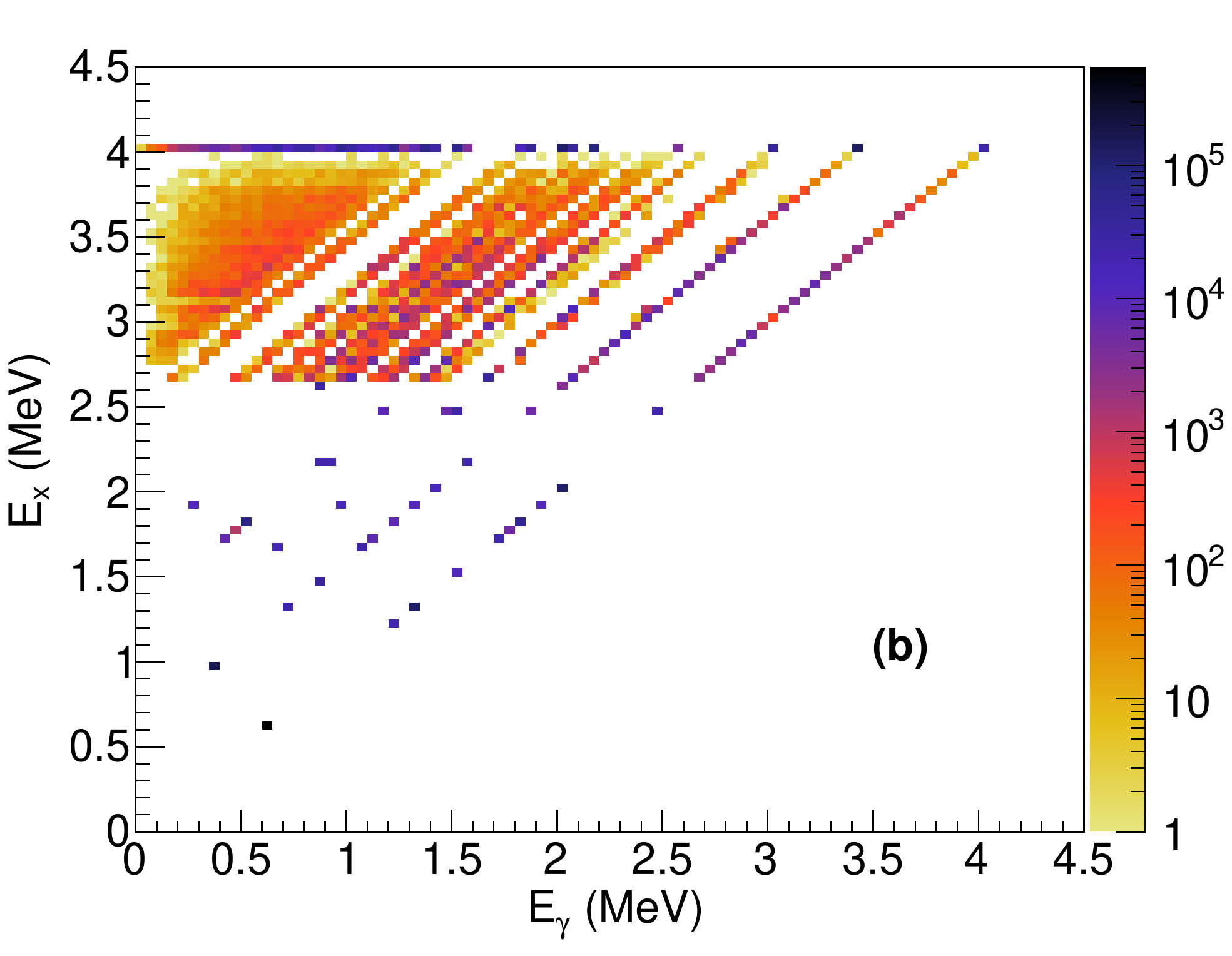}
}\hfill
\caption{(Color online) Capture cascade $\gamma$-ray emission intensities from 
the DICEBOX realizations that best match the data as a function of excitation 
energy and $\gamma$-ray energy.
These correspond to the red lines on the MSC plots in Figs.~\ref{fig:thermal_msc} 
and~\ref{fig:resonant_msc}.
(a) Cascade model for thermal neutron capture (b) Cascade model for 2.154 keV 
$^{136}\mathrm{Xe}$ neutron capture resonance.  Intensities (color scale) are 
expressed as transitions
per $10^{6}$ captures.}
\label{fig:cascades}
\end{figure*}

The decay scheme, represented as the relative intensities of emission as a 
function of
initial energy and $\gamma$-ray energy are, for both thermal and $E_n=2.154$~keV 
capture,
shown in Fig.~\ref{fig:cascades}. The intensities are given in 50 keV wide bins. 
Decay cascades corresponding to these schemes are included in the supplemental 
material.

As evident from Figs.~\ref{fig:thermal_msc} and~\ref{fig:resonant_msc}, 
simulations do not describe the spectra exactly, especially at
higher multiplicity. However, the discrepancies there are small compared to the 
entire 
intensity, accounting for only a few percent of all transitions. It should be 
noted that 
the number of counts from a cascade is given by $M_{\mathrm{cl}}$, 
so discrepancies in the MSC
histograms for higher multiplicities are exaggerated.

\subsection{Relative Cross Section}

Using the optimal nuclear realizations, we calculated the efficiency for 
detecting an event within $M_{\mathrm{cl}}=2-5$ and $E_\Sigma=3.625-4.225$~MeV. 
The efficiency for detecting a thermal (2.154~keV resonance) neutron capture 
within the selected $Q$~gate and multiplicity gate was 28.9\% (24.9\%). 

In general, the cross sections can be calculated as

\begin{equation}
\sigma(E_{n}) = \alpha\frac{N(E_{n})}{\epsilon(E_{n})\Phi(E_{n})},
\end{equation}
where $N$ is the number of captures passing selection cuts after background 
subtraction, $\epsilon$ is the efficiency for a capture to pass those 
selections, $\Phi$ is the neutron
flux, and $\alpha$ is a term containing the xenon gas density and other 
parameters which
are independent of neutron energy.
At thermal energy (25~meV) the cross section is near constant (to $\pm1\%$) 
within the measurement energy window.  At 
resonance energy, the cross section varies rapidly with $E_{n}$, so the integral 
of the cross section
over the resonance is the preferred way of reporting results.  Thus, the cross 
section
ratio between resonance and thermal captures is reported in units of inverse 
energy.
Reporting a ratio, rather than absolute cross sections, allows for considerable
reduction of systematic uncertainties, and avoids complications associated with
calibrating the absolute flux and efficiency.

The ratio of the cross section in the thermal window to the 2.154 keV resonance 
integral 
was found to be
4.10~\rpm~0.10~(stat.)~\rpm~0.24~(sys.)~$\mathrm{meV}^{-1}$. 
The thermal neutron energy
window was centered at 25~meV with a width of 1.3~meV and the resonance neutron
 energy window was chosen to be from 2094 to 2203 eV, which
encompasses the entire resonance within the neutron energy resolution of DANCE.  

The systematic uncertainty on the ratio comes from the quadrature sum of the 
flux ratio uncertainty (1.7\%), efficiency ratio 
uncertainty (3.3\%), 4027~keV direct transition uncertainty (0.5\%) and an 
additional uncertainty (4.4\%) which 
accounts for uncertainties in background subtraction. The background subtraction 
uncertainty was largely
determined through tests of the robustness of the measurement with different 
$Q$~gates.
The efficiency ratio uncertainty was computed by examining efficiency changes 
due to
possible energy mis-calibration, 
differences in efficiency between DAQ cards, crystal timing calibration, 
and simulation inaccuracy. 
The statistical uncertainty, comprising uncertainty in thermal, resonance,
and background counts, is 2.5\%. 

We also searched for peaks in the background-subtracted 
$M_{\mathrm{cl}}\ge 3$ event rate as a function of $E_{n}$.  
No additional resonances were observed for neutron energies between 
9 and 2154 eV.  In particular, we did not observe the 600 eV 
resonance included in the 
TENDL-2014 evaluation~\cite{tendl2014, KONING20122841}.

\section{\label{sec:abs_cross_section}Absolute Cross Section}

The ratio between thermal and resonance cross sections from the ENDF/B-VII.1 
evaluation~\cite{ENDF7xe136:2006:xxx} is 6.95~meV$^{-1}$, considerably different 
than
our measured ratio of 
4.10~\rpm~0.10~(stat.)~\rpm~0.24~(sys.)~$\mathrm{meV}^{-1}$.
After examining the calibration data available for the particular
DANCE detector configuration used, we found that we could determine a more precise
absolute thermal neutron cross section by using a
separate measurement of the resonance integral to calibrate our measurement.
One absolute cross section measurement of the \otsx 2.154~keV resonance has been 
reported by Macklin~\cite{Macklin}. 
Converting the resonance kernel value of
$30.1\pm1.5$~meV~\cite{Macklin} to a resonance integral yields 
$58.0\pm2.9$~b\,\,eV. 
Combining the relative cross section ratio from our analysis with this resonance 
integral gives us a value of 0.238~\rpm~0.019~b for the thermal cross section.

\begin{figure}[htb]
\includegraphics[width=\columnwidth]{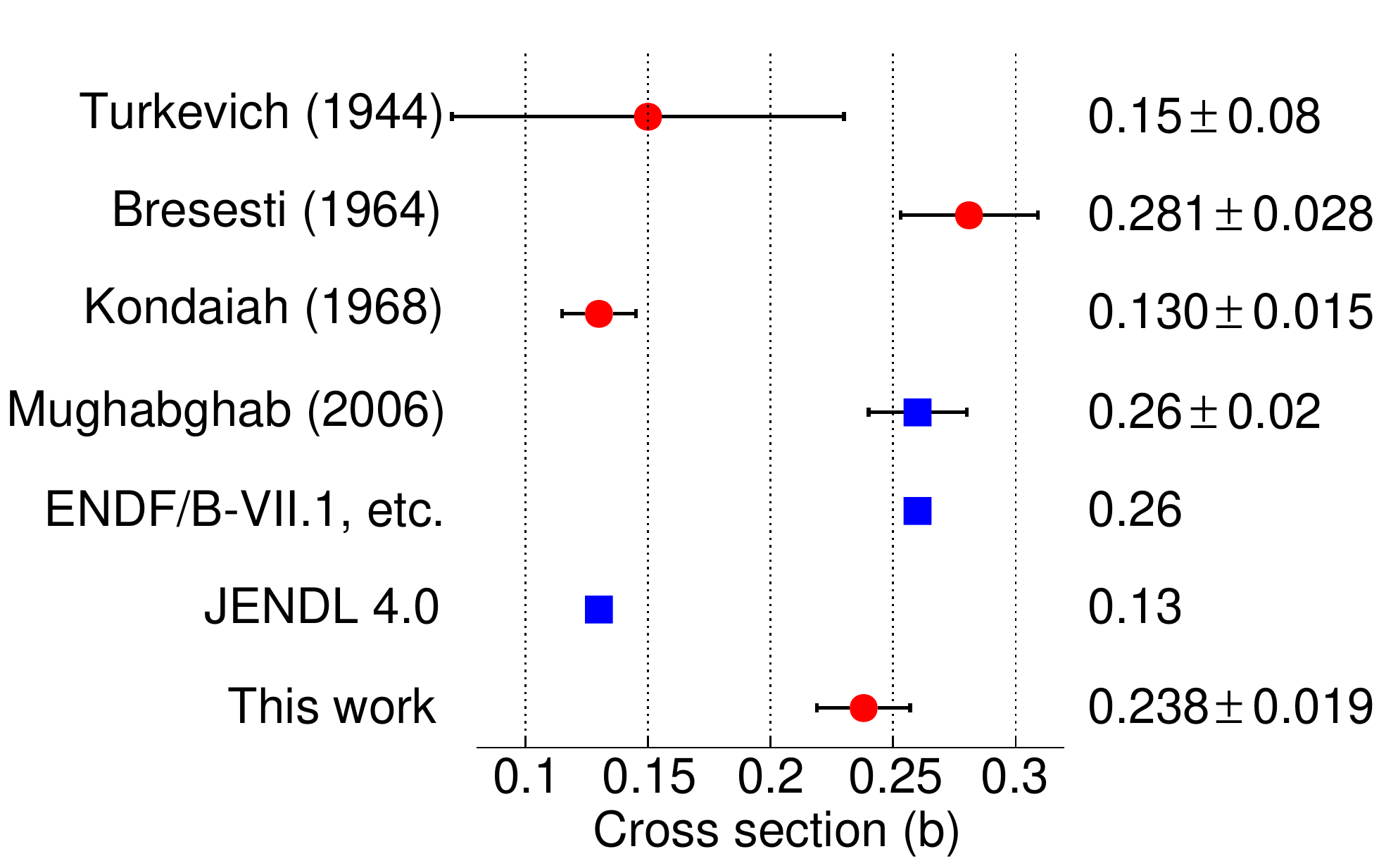}
\caption{(Color online) Comparison of various measurements and 
evaluations of the $^{136}\mathrm{Xe}(n,\gamma)$ cross 
section for thermal neutrons.  
Each row corresponds to a measurement (red circle) or evaluation (blue square),
and includes the cross section 
in both plot and text.  All cross sections
are in barns.
Measurements
by Macnamara~\textit{et al.} (1950)~\cite{PhysRev.80.296} and 
Eastwood~\textit{et al.} (1963)~\cite{eastwoodexfor} are less precise and are 
not included here.  Most modern evaluations, such as 
TENDL-2014~\cite{tendl2014, KONING20122841} and
JEFF-3.2~\cite{jeff32} have the same thermal neutron cross section as
 ENDF/B-VII.1~\cite{ENDF7xe136:2006:xxx}, 
so we do not list them  separately.  
JENDL-4.0~\cite{doi:10.1080/18811248.2011.9711675} is an exception
to this.  Information on the Turkevich~\textit{et al.} measurement comes
from Ref.~\cite{Bresesti19651175, Way_Haines_1947}.  Other results listed 
come from Ref.~\cite{mughabghab2006atlas, Bresesti19651175, Kondaiah1968329}.}
\label{fig:xstable}
\end{figure}

Past measurements of the thermal cross section
have considerable differences, and evaluated cross sections vary similarly.
A summary of thermal cross section measurements and evaluations is shown in
Figure~\ref{fig:xstable}.
Our
result favors the Bresesti \textit{et al.}~\cite{Bresesti19651175} measurement 
($0.281\pm0.028$~b) over the 
Kondaiah \textit{et al.}~\cite{Kondaiah1968329} measurement ($0.130\pm0.015$~b).  
Most recent evaluations~\cite{Pritychenko:2012ge}
give the thermal cross section as 0.26~b, consistent with our result, though 
JENDL-4.0~\cite{doi:10.1080/18811248.2011.9711675}
is an exception, favoring the Kondaiah measurement and giving 0.13~b.

\section{\label{sec:conclusions}Discussion}

The complete decay pattern from radiative neutron capture can only be obtained 
for light
nuclei using detectors with very good energy resolution (typically Ge).  More
complex nuclei, such as $^{137}\mathrm{Xe}$, have too many levels to obtain a 
perfect
cascade model.  Detectors with worse energy resolution but high granularity, 
such as
DANCE, can still provide valuable information about the cascades when 
simulations are
used to model the detector response and experimental spectra are compared with 
predicted models.

Fortunately, for purposes of modeling the cascades for use in neutrinoless 
double beta decay experiments such as EXO-200, it is not necessary to know the 
decay scheme with extremely high precision, and the approximation presented here 
is sufficient.
The Prussin~\textit{et al.} measurement of thermal neutron capture already 
produced a
capture cascade model with precisely measured energy levels.  The measurement 
presented here
features coincidence data not available in the previous measurement, and is used 
in
conjunction with the old results to produce a more refined capture cascade 
model. This may assist with mitigation 
of the $^{137}\mathrm{Xe}$ beta decay background in \znbb experiments.
The resonant capture model presented here is new.

The cross section measurement may help resolve discrepancies between earlier
measurements of thermal neutron capture on \otsx.  This can help guide
future evaluations,
which can in turn
allow for improved simulations of neutron transport in \znbb experiments. 

\vspace{3mm}

\section{Acknowledgements}

This work was partially funded by the Office of the Vice Provost of Research
at Indiana University-Bloomington through the Faculty Research Support
Program, and by the
US Department of Energy Grant No. DE-SC0012191.
This work has benefited from the use of the LANSCE facility at the Los Alamos 
National Laboratory.
This work was performed under the auspices of the U.S. Department of Energy by 
Los Alamos National Security, 
LLC, under Contract No. DE-AC52-06NA25396 and by Lawrence Livermore National 
Security, LLC, under Contract No. 
DE-AC52-07NA27344.
M.K. acknowledges the support of the Czech Science Foundation 
under Grant No. 13-07117S.

\bibliographystyle{./apsrev4-1}
\bibliography{dance-capture-crosssection}

\end{document}